\documentclass{emulateapj}

\shorttitle{Near-infrared thermal emission from TrES-3b} 
\shortauthors{Croll et al.}

\newcommand{\FpOverFStarPercentAbstractTrESThreeKLinear}{0.133}
\newcommand{\FpOverFStarPercentAbstractMinusTrESThreeKLinear}{0.016}
\newcommand{\FpOverFStarPercentAbstractPlusTrESThreeKLinear}{0.018}
\newcommand{\XSigmaTrESThreeKLinear}{8}
\newcommand{\FpOverFStarPercentAbstractThreeSigmaLimitTrESThreeKLinear}{0.185}

\newcommand{\PhaseAbstractTrESThreeKLinear}{0.5020}
\newcommand{\PhaseAbstractMinusTrESThreeKLinear}{0.0010}
\newcommand{\PhaseAbstractPlusTrESThreeKLinear}{0.0014}
\newcommand{\ECosOmegaTrESThreeKLinear}{0.0029}
\newcommand{\ECosOmegaPlusTrESThreeKLinear}{0.0022}
\newcommand{\ECosOmegaMinusTrESThreeKLinear}{0.0022}
\newcommand{\ECosOmegaAbsoluteThreeSigmaLimitTrESThreeKLinear}{0.0101}
\newcommand{\TBrightTrESThreeKLinear}{1731}
\newcommand{\TBrightPlusTrESThreeKLinear}{56}
\newcommand{\TBrightMinusTrESThreeKLinear}{60}
\newcommand{\TBrightThreeSigmaLimitTrESThreeKLinear}{1887}
\newcommand{\fReradiationTrESThreeKLinear}{0.303}
\newcommand{\fReradiationPlusTrESThreeKLinear}{0.042}
\newcommand{\fReradiationMinusTrESThreeKLinear}{0.040}
\newcommand{\fReradiationThreeSigmaLimitTrESThreeKLinear}{0.429}
\newcommand{\cOneTrESThreeKLinear}{0.00212}
\newcommand{\cOnePlusTrESThreeKLinear}{0.00017}
\newcommand{\cOneMinusTrESThreeKLinear}{0.00016}
\newcommand{\cTwoTrESThreeKLinear}{-0.025}
\newcommand{\cTwoPlusTrESThreeKLinear}{0.002}
\newcommand{\cTwoMinusTrESThreeKLinear}{0.002}

\newcommand{\TOffsetTrESThreeKLinear}{3.4}

\newcommand{\TOffsetPlusTrESThreeKLinear}{2.7}
\newcommand{\TOffsetMinusTrESThreeKLinear}{1.9}

\newcommand{\ChiTrESThreeKLinear}{1.044}
\newcommand{\ChiPlusTrESThreeKLinear}{0.006}
\newcommand{\ChiMinusTrESThreeKLinear}{0.001}
\newcommand{\JDOffsetONETrESThreeKLinear}{14985.9542}
\newcommand{\JDOffsetPlusONETrESThreeKLinear}{0.0019}
\newcommand{\JDOffsetMinusONETrESThreeKLinear}{0.0013}

\newcommand{\FpOverFStarPercentAbstractTrESThreeKLinearCyclicZero}{0.132}
\newcommand{\FpOverFStarPercentAbstractMinusTrESThreeKLinearCyclicZero}{0.010}
\newcommand{\FpOverFStarPercentAbstractPlusTrESThreeKLinearCyclicZero}{0.012}

\newcommand{\FpOverFStarPercentAbstractThreeSigmaLimitTrESThreeKLinearCyclicZero}{0.154}

\newcommand{\PhaseAbstractTrESThreeKLinearCyclicZero}{0.5021}
\newcommand{\PhaseAbstractMinusTrESThreeKLinearCyclicZero}{0.0012}
\newcommand{\PhaseAbstractPlusTrESThreeKLinearCyclicZero}{0.0010}
\newcommand{\ECosOmegaTrESThreeKLinearCyclicZero}{0.0030}
\newcommand{\ECosOmegaPlusTrESThreeKLinearCyclicZero}{0.0015}
\newcommand{\ECosOmegaMinusTrESThreeKLinearCyclicZero}{0.0015}

\newcommand{\TBrightTrESThreeKLinearCyclicZero}{1727}
\newcommand{\TBrightPlusTrESThreeKLinearCyclicZero}{39}
\newcommand{\TBrightMinusTrESThreeKLinearCyclicZero}{41}
\newcommand{\TBrightThreeSigmaLimitTrESThreeKLinearCyclicZero}{1799}
\newcommand{\fReradiationTrESThreeKLinearCyclicZero}{0.300}
\newcommand{\fReradiationPlusTrESThreeKLinearCyclicZero}{0.029}
\newcommand{\fReradiationMinusTrESThreeKLinearCyclicZero}{0.028}
\newcommand{\fReradiationThreeSigmaLimitTrESThreeKLinearCyclicZero}{0.354}
\newcommand{\cOneTrESThreeKLinearCyclicZero}{0.00211}
\newcommand{\cOnePlusTrESThreeKLinearCyclicZero}{0.00010}
\newcommand{\cOneMinusTrESThreeKLinearCyclicZero}{0.00009}
\newcommand{\cTwoTrESThreeKLinearCyclicZero}{-0.025}
\newcommand{\cTwoPlusTrESThreeKLinearCyclicZero}{0.001}
\newcommand{\cTwoMinusTrESThreeKLinearCyclicZero}{0.001}

\newcommand{\TOffsetTrESThreeKLinearCyclicZero}{3.6}

\newcommand{\TOffsetPlusTrESThreeKLinearCyclicZero}{1.8}
\newcommand{\TOffsetMinusTrESThreeKLinearCyclicZero}{2.2}

\newcommand{\ChiTrESThreeKLinearCyclicZero}{1.037}
\newcommand{\ChiPlusTrESThreeKLinearCyclicZero}{0.004}
\newcommand{\ChiMinusTrESThreeKLinearCyclicZero}{0.138}
\newcommand{\JDOffsetONETrESThreeKLinearCyclicZero}{14985.9543}
\newcommand{\JDOffsetPlusONETrESThreeKLinearCyclicZero}{0.0013}
\newcommand{\JDOffsetMinusONETrESThreeKLinearCyclicZero}{0.0015}

\newcommand{\FpOverFStarPercentAbstractTrESThreeH}{-0.002}
\newcommand{\FpOverFStarPercentAbstractMinusTrESThreeH}{0.019}
\newcommand{\FpOverFStarPercentAbstractPlusTrESThreeH}{0.015}

\newcommand{\FpOverFStarPercentAbstractThreeSigmaLimitTrESThreeH}{0.047}

\newcommand{\PhaseAbstractTrESThreeH}{0.5002}

\newcommand{\ECosOmegaTrESThreeH}{0.0000}

\newcommand{\TBrightThreeSigmaLimitTrESThreeH}{1635}

\newcommand{\fReradiationThreeSigmaLimitTrESThreeH}{0.242}
\newcommand{\cOneTrESThreeH}{0.00032}
\newcommand{\cOnePlusTrESThreeH}{0.00013}
\newcommand{\cOneMinusTrESThreeH}{0.00018}
\newcommand{\cTwoTrESThreeH}{-0.005}
\newcommand{\cTwoPlusTrESThreeH}{0.002}
\newcommand{\cTwoMinusTrESThreeH}{0.002}

\newcommand{\TOffsetTrESThreeH}{0.0}

\newcommand{\ChiTrESThreeH}{1.131}
\newcommand{\ChiPlusTrESThreeH}{0.006}
\newcommand{\ChiMinusTrESThreeH}{0.000}
\newcommand{\JDOffsetONETrESThreeH}{14989.8703}

\newcommand{\FpOverFStarPercentAbstractTrESThreeHCyclicZero}{0.011}
\newcommand{\FpOverFStarPercentAbstractMinusTrESThreeHCyclicZero}{0.040}
\newcommand{\FpOverFStarPercentAbstractPlusTrESThreeHCyclicZero}{0.019}

\newcommand{\FpOverFStarPercentAbstractThreeSigmaLimitTrESThreeHCyclicZero}{0.045}

\newcommand{\PhaseAbstractTrESThreeHCyclicZero}{0.5002}

\newcommand{\ECosOmegaTrESThreeHCyclicZero}{-0.0000}

\newcommand{\TBrightThreeSigmaLimitTrESThreeHCyclicZero}{1622}

\newcommand{\fReradiationThreeSigmaLimitTrESThreeHCyclicZero}{0.234}
\newcommand{\cOneTrESThreeHCyclicZero}{0.00010}
\newcommand{\cOnePlusTrESThreeHCyclicZero}{0.00055}
\newcommand{\cOneMinusTrESThreeHCyclicZero}{0.00003}
\newcommand{\cTwoTrESThreeHCyclicZero}{-0.004}
\newcommand{\cTwoPlusTrESThreeHCyclicZero}{0.002}
\newcommand{\cTwoMinusTrESThreeHCyclicZero}{0.006}

\newcommand{\TOffsetTrESThreeHCyclicZero}{-0.0}

\newcommand{\ChiTrESThreeHCyclicZero}{1.123}
\newcommand{\ChiPlusTrESThreeHCyclicZero}{0.024}
\newcommand{\ChiMinusTrESThreeHCyclicZero}{0.008}
\newcommand{\JDOffsetONETrESThreeHCyclicZero}{14989.8703}

\newcommand{\FpOverFStarPercentAbstractTrESThreeHCyclicOne}{-0.003}
\newcommand{\FpOverFStarPercentAbstractMinusTrESThreeHCyclicOne}{0.018}
\newcommand{\FpOverFStarPercentAbstractPlusTrESThreeHCyclicOne}{0.018}

\newcommand{\FpOverFStarPercentAbstractThreeSigmaLimitTrESThreeHCyclicOne}{0.051}

\newcommand{\PhaseAbstractTrESThreeHCyclicOne}{0.0000}

\newcommand{\ECosOmegaTrESThreeHCyclicOne}{-0.0003}

\newcommand{\TBrightThreeSigmaLimitTrESThreeHCyclicOne}{1658}

\newcommand{\fReradiationThreeSigmaLimitTrESThreeHCyclicOne}{0.255}
\newcommand{\cOneTrESThreeHCyclicOne}{0.00029}
\newcommand{\cOnePlusTrESThreeHCyclicOne}{0.00016}
\newcommand{\cOneMinusTrESThreeHCyclicOne}{0.00015}
\newcommand{\cTwoTrESThreeHCyclicOne}{-0.005}
\newcommand{\cTwoPlusTrESThreeHCyclicOne}{0.002}
\newcommand{\cTwoMinusTrESThreeHCyclicOne}{0.002}

\newcommand{\TOffsetTrESThreeHCyclicOne}{-0.4}

\newcommand{\ChiTrESThreeHCyclicOne}{1.128}
\newcommand{\ChiPlusTrESThreeHCyclicOne}{0.002}
\newcommand{\ChiMinusTrESThreeHCyclicOne}{0.004}
\newcommand{\JDOffsetONETrESThreeHCyclicOne}{14989.8701}

\newcommand{\BlackbodyOneChi}{19.0}	%f=0.25
\newcommand{\BlackbodyTwoChi}{16.6}	%f=0.295
\newcommand{\FortneyOneChi}{29.1}	%TiO/VO f=0.37
\newcommand{\FortneyTwoChi}{18.4}	%TiO/VO f=0.29
\newcommand{\FortneyThreeChi}{5.6}	%no TiO/VO f=0.37
\newcommand{\FortneyFourChi}{16.1}	%no TiO/VO f=0.29

\newcommand{\BlackbodyOneChiNoH}{26.8}	%f=0.25
\newcommand{\BlackbodyTwoChiNoH}{28.7}	%f=0.295
\newcommand{\FortneyOneChiNoH}{39.7}	%TiO/VO f=0.37
\newcommand{\FortneyTwoChiNoH}{25.6}	%TiO/VO f=0.29
\newcommand{\FortneyThreeChiNoH}{21.1}	%no TiO/VO f=0.37
\newcommand{\FortneyFourChiNoH}{25.7}	%no TiO/VO f=0.29

\newcommand{\fReradiationTrESThreeALL}{0.301}
\newcommand{\fReradiationPlusTrESThreeALL}{0.026}
\newcommand{\fReradiationMinusTrESThreeALL}{0.025}
\newcommand{\XPointXXKband}{13.8$\times$10$^{-3}$}
\newcommand{\YPointYYKband}{1.60$\times$10$^{-3}$}
\newcommand{\HowManyK}{9}
\newcommand{\XPointXXHband}{5.9$\times$10$^{-3}$}
\newcommand{\YPointYYHband}{0.84$\times$10$^{-3}$}
\newcommand{\HowManyH}{10}

\newcommand{\HowManySigmaTrESTwo}{5}

\begin{document}

\title{Near-infrared thermal emission from TrES-3b: a Ks-band detection and an H-band upper limit
on the depth of the Secondary Eclipse\altaffilmark{*}}
\author{Bryce Croll\altaffilmark{1},
Ray Jayawardhana\altaffilmark{1}, 
Jonathan J. Fortney\altaffilmark{2},
David Lafreni\`ere\altaffilmark{3},
Loic Albert\altaffilmark{4}
%Marten van Kerkwijk \altaffilmark{1}
}

\altaffiltext{1}{Department of Astronomy and Astrophysics, University of Toronto, 50 St. George Street, Toronto, ON 
M5S 3H4, Canada;
croll@astro.utoronto.ca}

\altaffiltext{2}{Department of Astronomy and Astrophysics, University of California, Santa Cruz, CA, 95064}

\altaffiltext{3}{D\'epartement de physique, Universit\'e de Montr\'eal, C.P.
6128 Succ. Centre-Ville, Montr\'eal, QC, H3C 3J7, Canada}

\altaffiltext{4}{Canada-France-Hawaii Telescope Corporation, 65-1238 Mamalahoa Highway,
Kamuela, HI 96743.}

\altaffiltext{*}{Based on observations obtained with WIRCam, a joint project of CFHT, Taiwan, Korea, Canada, France, at the Canada-France-Hawaii Telescope (CFHT) which is operated by the National Research Council (NRC) of Canada, the Institute National des Sciences de l'Univers of the Centre National de la Recherche Scientifique of France, and the University of Hawaii.}

\begin{abstract}

We present H and Ks-band photometry bracketing 
the secondary eclipse of the hot Jupiter TrES-3b
using the Wide-field Infrared Camera on the Canada-France-Hawaii Telescope.
We detect the secondary eclipse of TrES-3b with a depth of 
\FpOverFStarPercentAbstractTrESThreeKLinear$^{+\FpOverFStarPercentAbstractPlusTrESThreeKLinear}_{-\FpOverFStarPercentAbstractMinusTrESThreeKLinear}$\% in Ks-band
(\XSigmaTrESThreeKLinear$\sigma$) - a result in
sharp contrast to the eclipse depth reported by de Mooij \& Snellen.
We do not detect its 
thermal emission in H-band, but place a 3$\sigma$ limit on the depth 
of the secondary 
eclipse in this band of \FpOverFStarPercentAbstractThreeSigmaLimitTrESThreeHCyclicOne\%.
A secondary eclipse
of this depth in Ks requires very efficient day-to-nightside redistribution of heat and nearly 
isotropic reradiation,
% from an atmosphere that does not harbour a temperature inversion,
a conclusion
that is in agreement with longer wavelength, mid-infrared Spitzer observations.
Our 3$\sigma$ upper-limit on the depth of our H-band secondary
eclipse also argues for very efficient redistribution of heat and suggests that the atmospheric layer probed by these 
observations may be well homogenized. However, our H-band upper limit is so constraining that it suggests the possibility of a temperature inversion at depth, or an 
absorbing molecule, such as methane,
that further depresses the emitted flux at this wavelength.
The combination of our near-infrared measurements and those obtained with Spitzer suggest that TrES-3b displays a near
isothermal dayside atmospheric temperature structure, whose spectrum is well approximated by a blackbody. 
We emphasize that our strict H-band limit is in stark disagreement with the best-fit atmospheric model that results from 
longer wavelength observations only,
thus highlighting the importance of near-infrared observations at multiple wavelengths in addition to those returned 
by Spitzer in the mid-infrared
to facilitate a comprehensive understanding of the energy budgets of transiting exoplanets.
\end{abstract}


\section{Introduction}

Near-infrared secondary eclipse observations of hot Jupiters from 
the ground is a relatively new field.
After a series of non-detections and increasingly more sensitive upper limits using photometry
(\citealt{Snellen05}; \citealt{SnellenCovino07}; \citealt{Deming07}) and spectroscopy (\citealt{Richardson03}; \citealt{Knutson07}),
the field
has been reinvigorated by a series of successful ground-based, photometric
detections.
Examples include:
a $\sim$6$\sigma$ detection in
Ks-band of TrES-3b using the William Herschel Telescope (WHT; \citealt{deMooij09}),
a $\sim$4$\sigma$ detection in z'-band emission
of OGLE-TR-56b using Magellan and the Very Large Telescope (VLT; \citealt{Sing09}), 
a $\sim$5$\sigma$ detection at $\sim$2.1 $\mu m$ with the VLT \citep{Gillon09}, a $\sim$8$\sigma$ detection in 
the Ks-band of CoRoT-1b using the Apache Point observatory (APO; \citealt{Rogers09}), and
a $\sim$5$\sigma$ detection of WASP-12b's z'-band emission also using the APO \citep{LopezMorales10}. From our own
program we were able to report a $\sim$\HowManySigmaTrESTwo$\sigma$ detection of thermal emission from the hot Jupiter TrES-2b
in Ks-band \citep{CrollTrESTwo} using the Wide-field InfraRed Camera  (WIRCam) on 
the Canada-France-Hawaii Telescope (CFHT).

Near-infrared measurements of the thermal emission of hot Jupiters are crucial to our understanding
of the dynamics and radiative transfer in the atmospheres of these exotic worlds as these measurements sample
their blackbody peaks.
Such near-infrared measurements, when combined with secondary eclipse
detections longwards of 3$\mu m$
with Spitzer, enable us to characterize these planets'
pressure-temperature profiles and better understand their energy budgets.
Specifically they facilitate an estimate of the 
bolometric luminosity of these planets' dayside emission \citep{Barman08},
leading to a more complete understanding of how the planets reradiate the incoming stellar flux and advect
this heat from the day to nightside at various depths and pressures.

One of the most favourable targets for ground-based measurements is the 
transiting hot Jupiter TrES-3b. It circles
a G-type star in a $\sim$31 hour orbit \citep{ODonovan07}.
It is exposed to relatively high stellar insolation, with an incident flux of $1.7$$\times$$10^{9}$ $erg$$s^{-1}$$cm^{-2}$, 
and is thus
a member of the hottest and mostly highly irradiated class (pM-class) of hot Jupiters according
to the \citet{Fortney08} theory. Its high equilibrium
temperature ($T_{EQ}$$\sim$1650 $K$; assuming isotropic reradiation, and a zero Bond albedo)
in combination with its relatively favourable
planet-to-star radius ratio ($R_{P}/R_{*}$$\sim$0.166; \citealt{Sozzetti09}),
makes it a compelling target
for thermal emission measurements. 

 Thermal emission from this target has already been measured with Spitzer in the four IRAC \citep{Fazio04} channels
\citep{Fressin09}.
Their best-fit eclipes are consistent with a circular orbit, and \citet{Fressin09} place a 3$\sigma$ limit on the eccentricity, $e$, and
argument of periastron, $\omega$, of $|$$e$cos$\omega$$|$ $<$ 0.0056.
Despite the high incident stellar irradation for this target, their secondary
eclipse measurements are best-fit with an atmospheric model that efficiently redistributes heat.
Also, as the 4.5 $\mu m$ eclipse depth is less than the 3.6 $\mu m$ depth,
this suggests that this planet does not harbour a temperature inversion, as we are seeing water in absorption
rather than emission at 4.5 $\mu m$.
This is surprising,
because highly irradiated hot Jupiters, such as TrES-3b,
were expected to experience temperature inversions 
due to absorption of the incoming stellar flux from gaseous TiO/VO in a hot stratosphere \citep{Hubeny03,Fortney06,Burrows07,Fortney08}.

In the near-infrared, \citet{deMooij09}
have reported a detection of TrES-3b's 
thermal emission in Ks-band using the LIRIS instrument on the 
WHT. The depth of their best-fit secondary eclipse 
was 0.241 $\pm$ 0.043\%, and this result 
argued in favour of very bright dayside emission and very inefficient redistribution of heat to the nightside
of this planet, in sharp disagreement to the Spitzer results from longer wavelengths.
However, the authors noted residual 
systematic noise during the ingress of the secondary eclipse that resulted in a deep,
slightly eccentric best-fit eclipse.
For these reasons
we felt follow-up observations were warranted to confirm their measured eclipse depth.

A bright secondary eclipse in the near-infrared is reasonable,
as simplified one-dimensional, radiative transfer models \citep{Hubeny03} suggest that one can
expect hot Jupiters without temperature inversions
to display increased thermal emission in the near-infrared. 
This makes intuitive sense as the decreased output 
in the mid-infrared allows the planet to shine more brightly at shorter wavelengths.
On the other hand, near-infrared observations are expected to probe atmospheric layers that are more homogenized
than the layers probed by longer wavelength observations. This is because the JHK
near-infrared spectral bands occur at minima in the opacity of water where one should be able to see deeper into
a planet's atmosphere than one can see in the mid-infrared with \emph{Spitzer} \citep{Seager05,Fortney08,Burrows08}.
JHK-band observations should then probe higher pressure ($P$) atmospheric layers. At higher pressures it is expected
that the radiative time-scale (how quickly the planet
reradiates the incident stellar flux; $\tau_{rad}$)
will become of similar order to the advective timescale (how quickly the planet advects the heat to the nightside of the planet; $\tau_{adv}$)
leading to a more homogenized atmospheric layer \citep{Seager05,Fortney08}. 
The reradiative timescale is thought to be proportional to pressure:
$\tau_{rad}$ $\sim$ $\frac{P c_{P}}{g 4 \sigma T^3}$ \citep{ShowmanGuillot02}, where $T$ is the temperature,
$c_{P}$ is the specific heat capacity, 
$\sigma$ is the Stefan-Boltzmann constant, and $g$ is the gravitational
acceleration of the planet. The advective timescale, on the other hand,
is thought to be approximated by the radius of the planet, $R_{P}$, divided by the horizontal windspeed, $U$: $\tau_{adv}$ $\sim$ $R_{P}$/U \citep{ShowmanGuillot02}.
For these reasons, as one probes higher pressure atmospheric layers, $\tau_{rad}$ should increase,
and become of similar order to $\tau_{adv}$; this has been confirmed by 3D models \citep{Showman09}. 
For the near-infrared JHK bands we therefore may expect more efficient heat redistribution
and that the near-infrared emission may be depressed, as the depths we probe may be more homogenized than 
the upper atmospheres of these planets.  This behavior is difficult to capture for a 1D
radiative-equilibrium model that assumes an average day-side temperature  or planet-wide conditions. 
For these reasons near-infrared
observations are crucial to inform our understanding of the underlying physics that govern hot Jupiter
radiative transfer and atmospheric dynamics.

% Simple estimates of radiative and advective timescales argue for the planet being more homogenized at depth
% , and this can be tested with observations, such as these, in the near-infrared.
% so that at wavelengths where low pressures are probed (such as the mid-infrared, where water opacity is generally high)
% the planet's emission is bright, an effect further enhanced if the upper atmosphere is hot due to a temperature inversion. 

Here we present
observations bracketing TrES-3b's secondary eclipse using WIRCam on CFHT.
We report a \XSigmaTrESThreeKLinear $\sigma$ detection of its Ks-band thermal emission of 
\FpOverFStarPercentAbstractTrESThreeKLinear$^{+\FpOverFStarPercentAbstractPlusTrESThreeKLinear}_{-\FpOverFStarPercentAbstractMinusTrESThreeKLinear}$\%
and place a 3$\sigma$ upper limit on its thermal emission in H-band of 
\FpOverFStarPercentAbstractThreeSigmaLimitTrESThreeHCyclicOne \%. 
We do not find that this planet radiates brightly in the near-infrared, as our Ks-band measurement and our H-band upper limit
argue in favour of very efficient day-to-nightside redistribution and nearly isotropic reradiation of heat. 
These results are in clear contrast to the \citet{deMooij09} result, although we explain that under different assumptions their
result is consistent with our own.
Our H-band limit is so constraining that it suggests that the layer of the atmosphere
probed by this wavelength is very well homogenized and that there may be a temperature
inversion deep in this planet's atmosphere,
or that the emitted flux is depressed at this wavelength
due to a wide absorption band near 1.6 $\mu m$. 

\section{Observations and data reduction}
\label{SecReduction}

\begin{figure*}
\centering
\includegraphics[scale=0.55,angle=90]{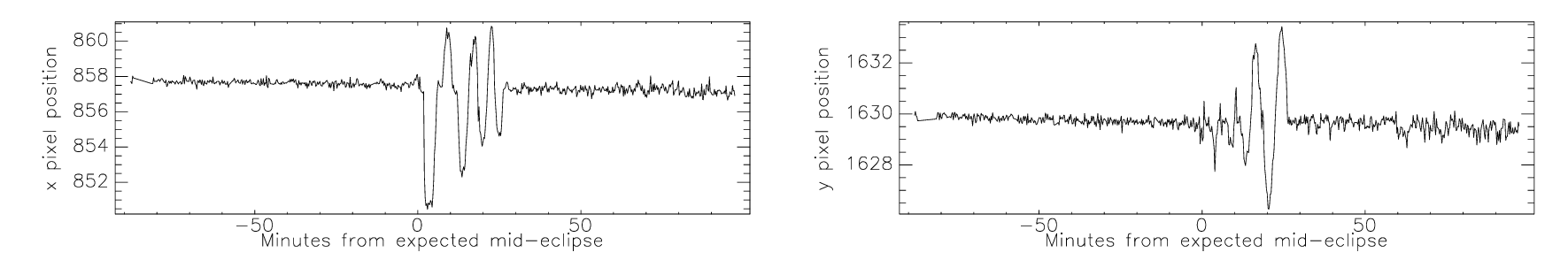}
\includegraphics[scale=0.55,angle=90]{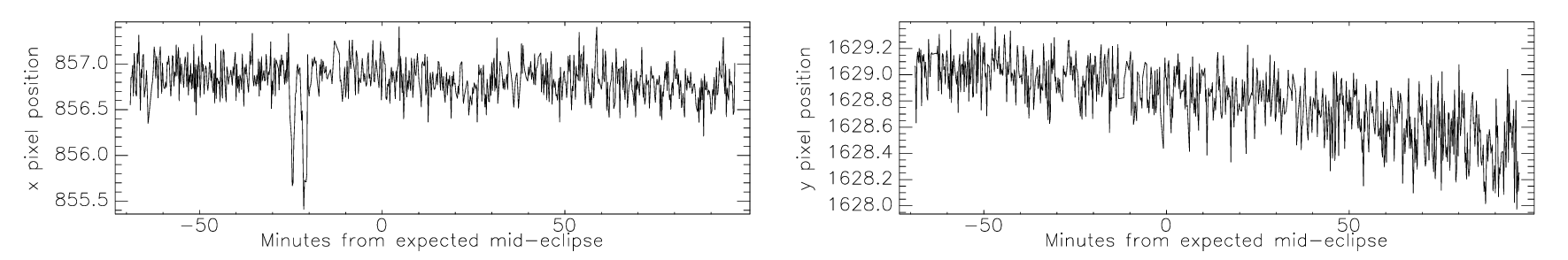}
\caption{	The x and y position of the centroid of the PSF of the target star, TrES-3, with time for our Ks-band photometry (top panels)
		and our H-band photometry (bottom panels).
	}
\label{FigDrifts}
\end{figure*}

\begin{figure*}
\centering
\includegraphics[scale=0.40,angle=90]{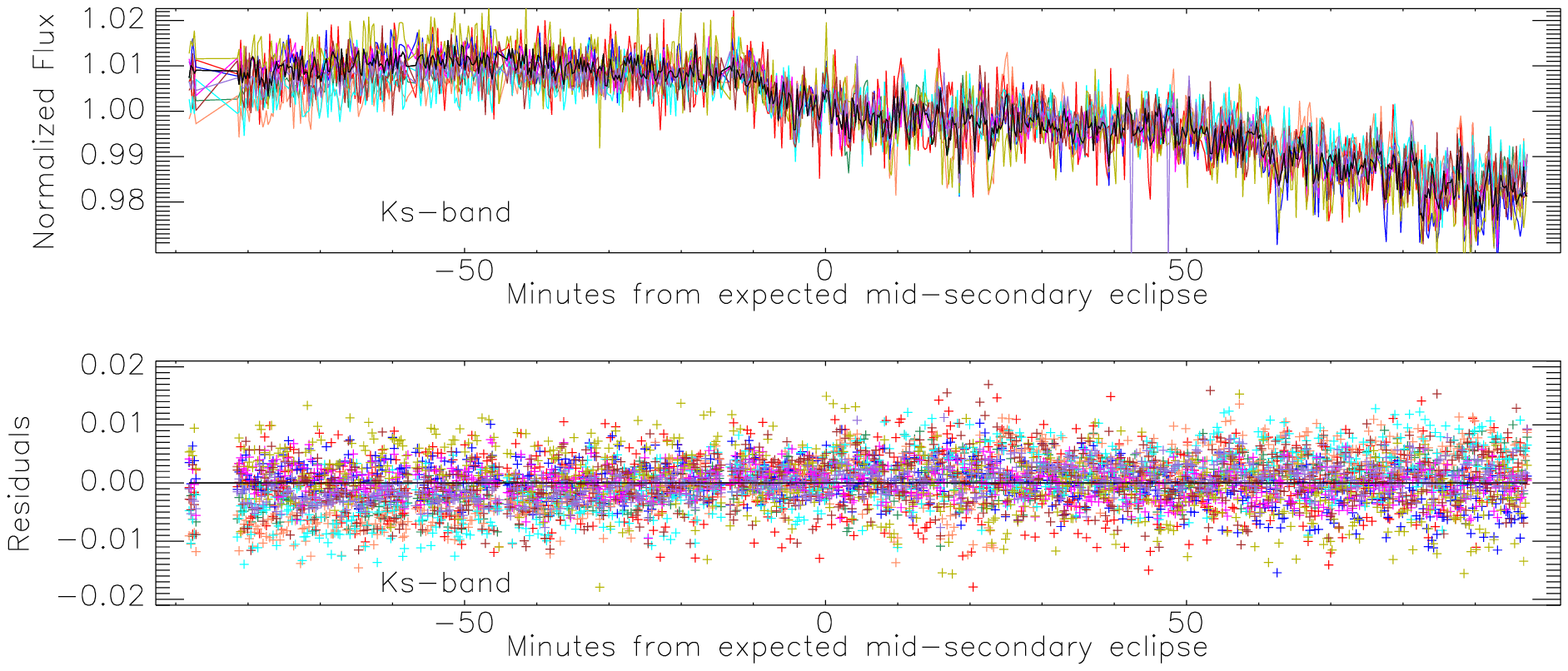}
\includegraphics[scale=0.40,angle=90]{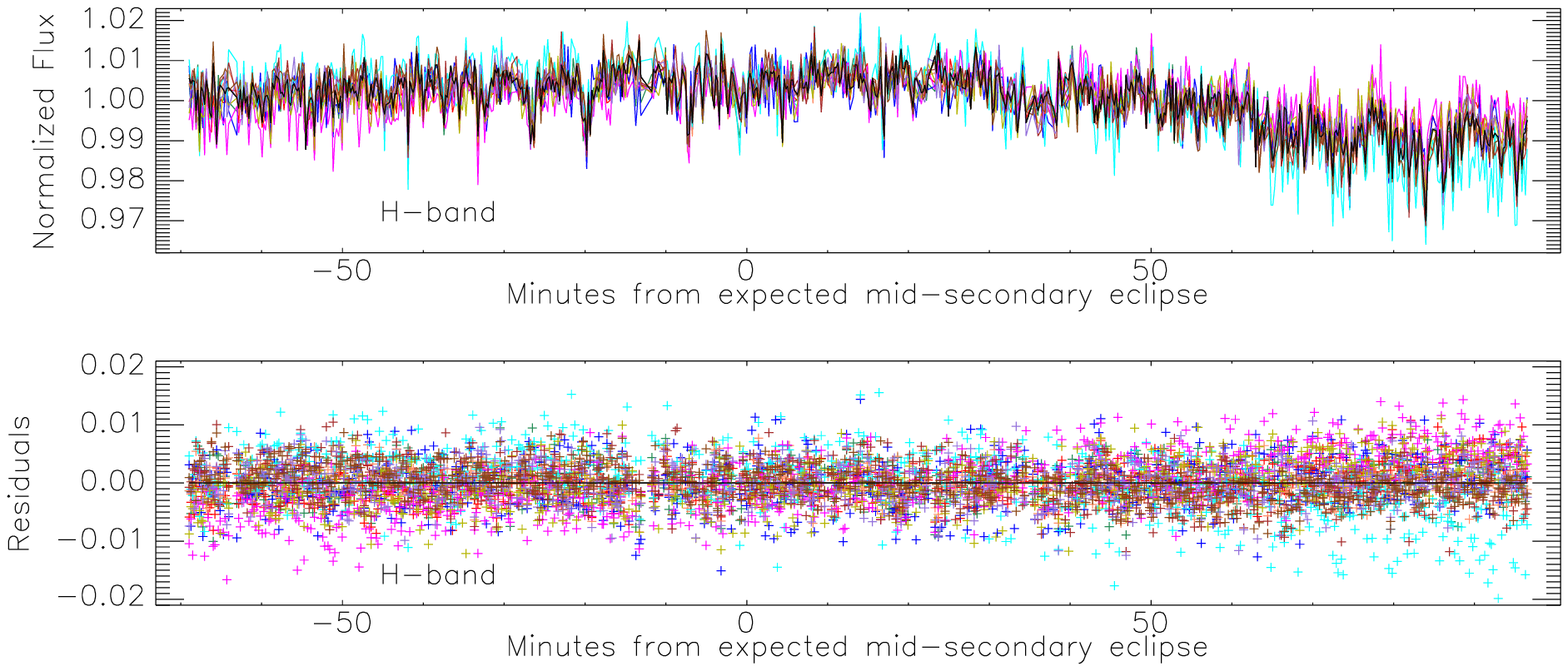}
\caption{	Top panels: The flux from the target star (black) and the reference stars (various colours)
		that are used to calibrate the flux of TrES-3b for our Ks-band photometry (left)  and our H-band photometry (right).
		Bottom panels: The residuals from the normalized flux of the target star 
		of the normalized flux of the reference stars for the Ks-band (left) and H-band photometry (right).
		%CHANGE_HERE if necessary
		%%Given the large number of reference stars for our Ks-band photometry (left)
		%%we bin the data every three points in time to ensure clarity.
	}
\label{FigTrES3bRefStars}
\end{figure*}

 We observed the secondary eclipse of TrES-3 ($K$=10.608, $H$=10.655)
on two occasions with WIRCam on CFHT \citep{Puget04}.
On 2009 June 3 we observed TrES-3
using a Ks-band filter and on 2009 June 6 we observed it again using an H-band filter.
Both observations were taken under
photometric conditions and lasted for $\sim$3.1 hours for our Ks observations
and $\sim$2.8 hours for our H-band observations,
evenly bracketing the predicted secondary eclipse.
On both occasions numerous reference stars
were also observed in the 21x21 arcmin field of view of WIRCam.
The telescope was defocused to 1.0mm, resulting in the
flux of our target star being spread over a ring 13 pixels in diameter (4\arcsec) on our array.
We used ``stare'' mode on CFHT where the target star is observed continuously without dithering for the duration of the observations.

We used 5-second exposures for our Ks-band observations. The effective duty cycle
after accounting for readout and for saving exposures was 33\%. 
Following the observations we noted significant drifts in the centroid of the
stellar point-spread-function (PSF) of TrES-3 as well as other stars on the chip (Figure \ref{FigDrifts} top panels).

For our H-band observations, and subsequent observations in our program, to counteract
these drifts we initiated a corrective guiding ``bump'' before every image cube to ensure 
that our target star fell as often as possible on or near the original pixel. 
Following this corrective ``bump'' the drifts on the chip were significantly reduced 
(Figure \ref{FigDrifts} bottom panels). 
To counteract
the extra overheads that this ``bump'' induced, we observed in ``cubes'' of multiple images in 
each FITS files. 
Thus, for 
our H-band observations we obtained data-cubes each containing 
twelve 5-second exposures.
The effective duty cycle
after accounting for readout and for saving exposures was 43\%.

%CHANGE_HERE - possibly change due to reduction changes
For both sets of data the images were reduced and aperture photometry was performed
on our target star and all unsaturated,
reasonably bright reference stars on the WIRCam array
as discussed in \citet{CrollTrESTwo}.
Exceptions include that we do not modify the shape of the annulus used to calculate the sky
aperture, and that we do not correct the flux of our targets for the x or y position of
the centroid of the stellar PSF. We used an aperture with a radius of 11 pixels for our Ks-band photometry, and an aperture
with a radius of 10.5 pixels for our H-band photometry. To estimate the residual background flux for both
sets of photometry we used an annulus with an inner
radius of 18 pixels and an outer radius of 28 pixels.
We tested larger and smaller apertures in increments of 0.5 pixels, and confirmed that these sizes of apertures returned optimal
photometry.

For our H-band photometry pixels in the annulus produced by our defocused target star 
occassionally saturated, and were removed in the preprocessing
step; similar saturation issues were noted with some of our reference stars as well. It was difficult
to account for the discrepancy in flux that resulted from these saturated pixels at the precision required for these observations;
thus observations in which a pixel near our target star were saturated 
were excluded from the resulting analysis. 
62 of our 718 H-band observations ($\sim$8.6\% of the total) were cut as a result of this step.
Saturation was not an issue in the 740 exposures for our Ks-band photometry.

%The flux of our target and reference stars were also observed to display a linear
%correlations with the x or y position on the centroid
%of the stellar point-spread-function (PSF) on the chip.
%For both sets of observations the
%stellar PSF still displayed high frequency 
%shifts in position (Figure \ref{FigDrifts} top panels).
%As the frequecy of these shifts was high it effectively ensured
%that any trend with the flux of the star was instrumental in origin.
%Thus these linear trends were removed
%from the data for both the target and reference stars 
%as displayed in Figure \ref{FigDrifts} bottom panels.

\begin{figure}
\centering
\includegraphics[scale=0.45, angle = 270]{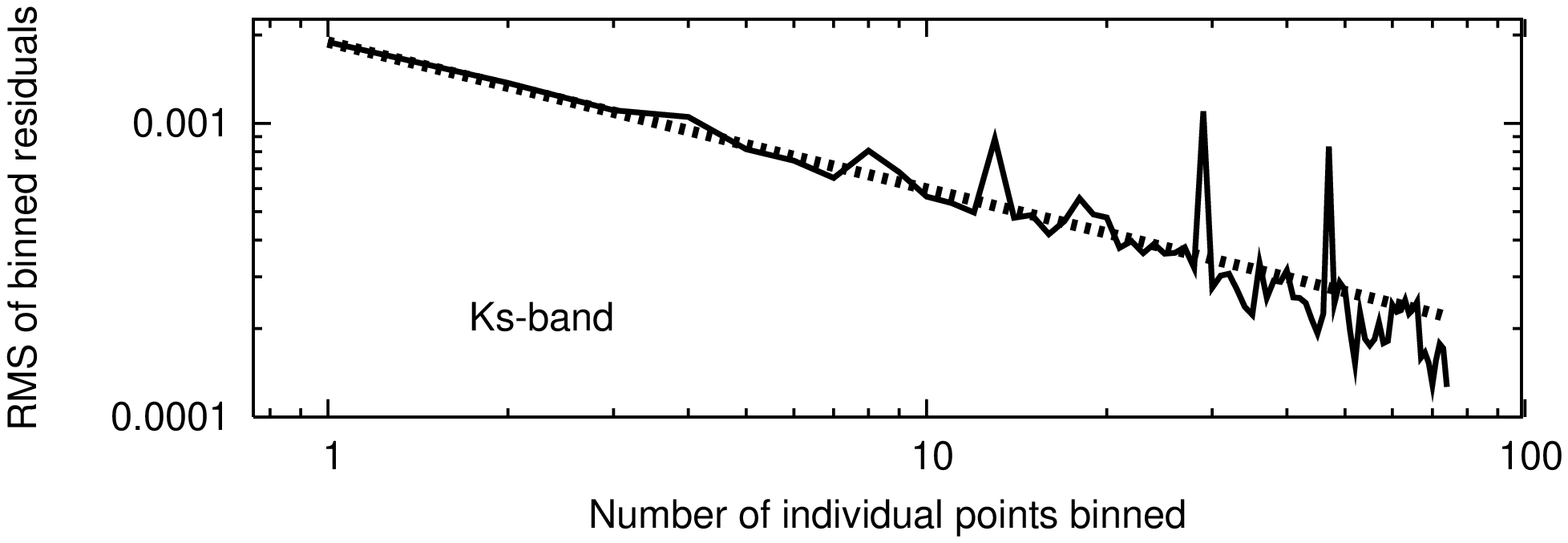}
\includegraphics[scale=0.45, angle = 270]{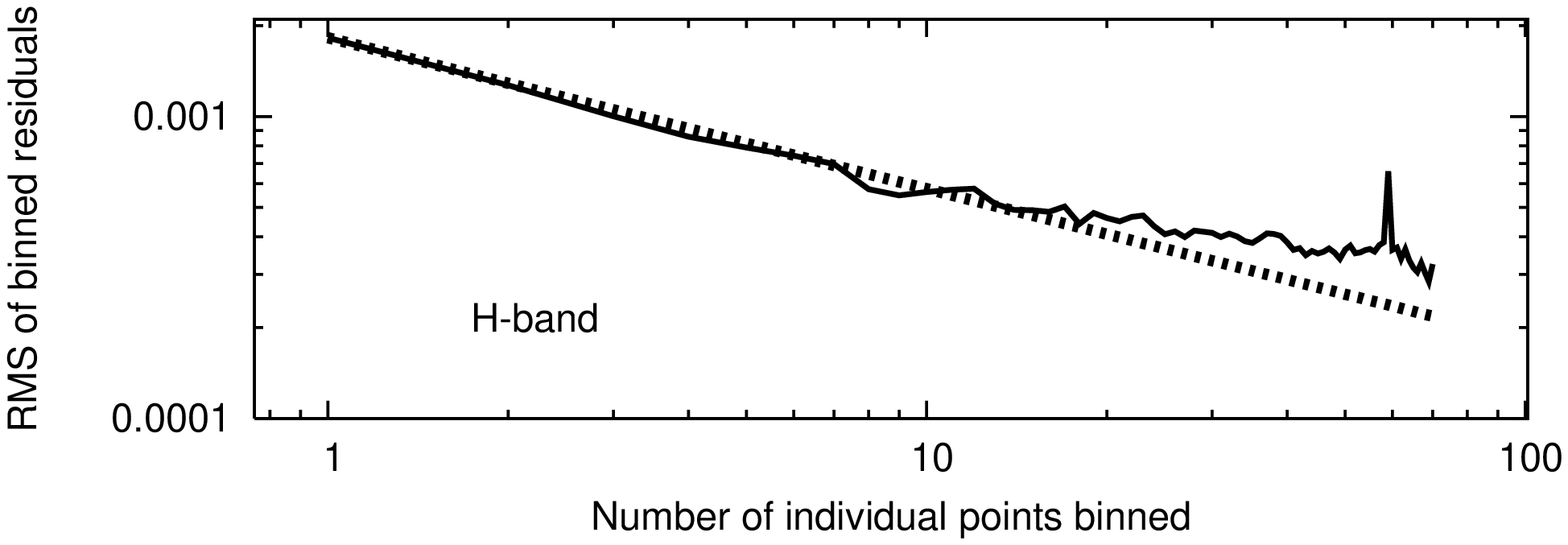}
\caption{	The root-mean-square of our out-of-eclipse photometry (solid line) for various number of binned points
		following the various corrections documented in $\S$\ref{SecReduction}
		for our Ks-band photometry (top) and our H-band photometry (bottom).
		In both cases the dashed line displays the
		one over the square-root of the bin-size expectation for gaussian noise.
		}
\label{FigPoisson}
\end{figure}

As with our previous near-infrared CFHT/WIRCam photometry \citep{CrollTrESTwo},
the resulting light curves displayed significant, systematic variations in intensity 
(see the top panels of Figure \ref{FigTrES3bRefStars}), possibly due to changes in
atmospheric transmission, seeing and airmass, guiding errors and/or other effects.
These variations in the flux of our target star were then corrected by normalizing the flux of the 
target star by
\HowManyK \ reference stars in Ks-band and \HowManyH \ in H-band using the method
discussed in \citet{CrollTrESTwo}.
This method entails using the reference
stars that showed the smallest deviation from the target star outside of the expected secondary 
eclipse to correct the flux of our target star.
As a result
of these corrections (and the removal of a linear-trend with time for the sake of this comparison only),
the point-to-point scatter of our data outside of occultation
improved from a root-mean-square 
of \XPointXXKband \ to \YPointYYKband \ in Ks and \XPointXXHband \ to \YPointYYHband \ in H 
per 60 seconds (Figure \ref{FigTrES3bRefStars}). 
We should note that we are still well above the predicted photon noise RMS limit
of 3.7$\times$10$^{-4}$ for Ks-band, and
2.7$\times$10$^{-4}$ for H-band per 60 seconds. 
We set the
uncertainty on our measurements as the RMS
of the out of eclipse photometry after the removal of a
linear-trend with time.
We also bin our out-of-eclipse photometry
following the above reduction and compare it to the one-over-the square-root of the number of binned points expectation for gaussian noise.
Although our Ks-band data scales down near this limit,
our H-band data displays systematics that result in the data
scaling down marginally above this limit (Figure \ref{FigPoisson}).
This suggests
that there is an extra systematic in our H-band photometry that merits further investigation.

\section{Analysis}

\begin{figure*}
\centering
\includegraphics[scale=0.30,angle=270]{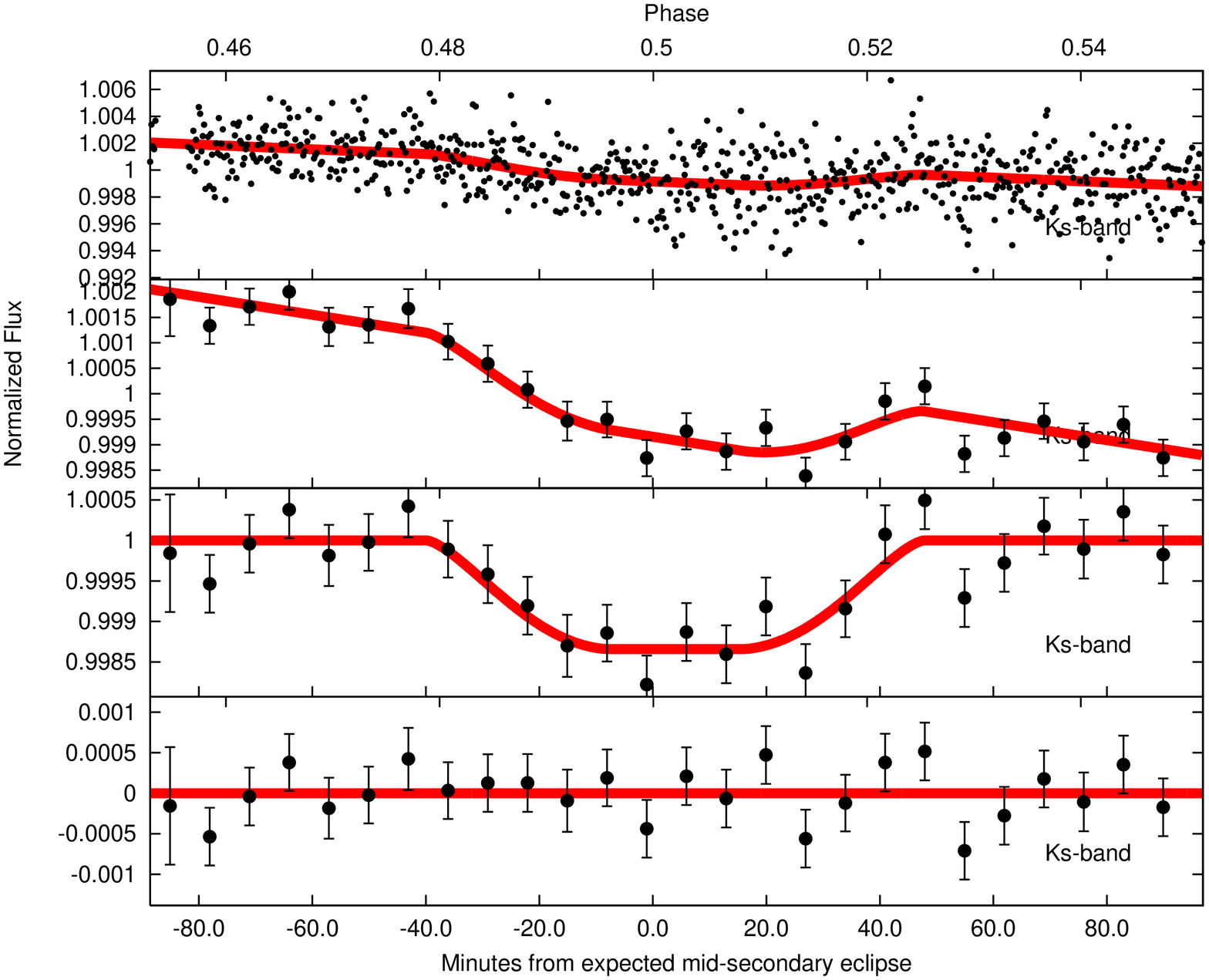}
\includegraphics[scale=0.30,angle=270]{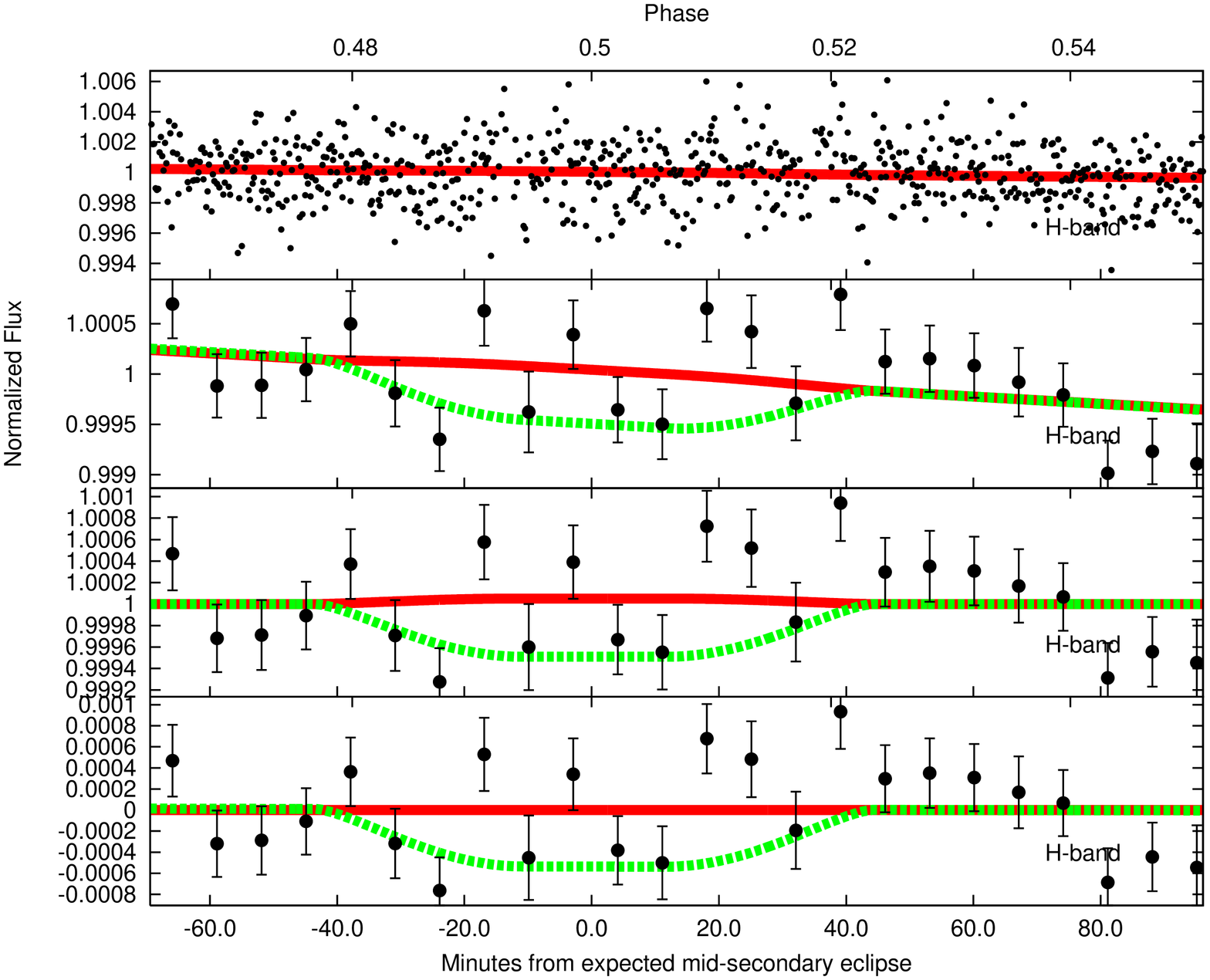}
\caption{	
		WIRCam/CFHT photometry bracketing the secondary eclipse of TrES-3b
		in Ks-band (left) and H-band (right).
		The top panels show the unbinned lightcurves, the panel that is second from the top shows the lightcurves
		with the data binned every 7.0 minutes.
		The panel that is the second
		from the bottom shows the binned data after the subtraction of
		the best-fit background trends, $B_f$ (the slope), while 
		the bottom panels show the binned residuals from the best-fit model of each eclipse.
		In each one of the panels the best-fit secondary eclipse and background trend, $B_f$,
		is shown with the red line.
		For our H-band photometry at right we also display the depth of the secondary eclipse that 
		we are able to rule out at 3$\sigma$ (green dotted-line). As displayed in Table \ref{TableParams}
		the best-fit secondary eclipse in H-band (right panel) has a small
		negative depth (thus representing an unphysical brightening).
	}
\label{FigObservations}
\end{figure*}

\begin{figure}
\centering
\includegraphics[scale=0.65,angle=270]{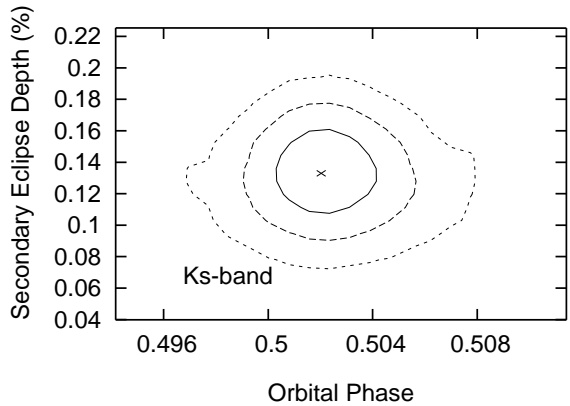}
\caption{	The 68.3\% (1$\sigma$; solid-line), 95.5\% (2$\sigma$; dashed-line)
		and 99.7\% (3$\sigma$; short dashed-line) credible regions
		from our MCMC analysis for our Ks-band photometry comparing the secondary eclipse depth, $\Delta F$, and the best-fit phase, $\phi$. The ``x'' in the middle of the plot denotes the best-fit point from our 
		MCMC analysis.}
\label{FigContour}
\end{figure}

%emulateApJ change
\begin{deluxetable*}{cccccc}
%\begin{deluxetable}{cccccc}
\tabletypesize{\tiny}
\tablecaption{Best-fit secondary eclipse parameters}
\tablehead{
\colhead{Parameter} 	& \colhead{MCMC Ks-band } 	& \colhead{``Residual permutation''}	& \colhead{MCMC H-band} 	& \colhead{``Residual permutation''}	& \colhead{Bootstrap}		\\
\colhead{}		& \colhead{Solution}		& \colhead{Ks-band solution}		& \colhead{Solution}		& \colhead{H-band solution}		& \colhead{H-band solution}	\\
}
\startdata
reduced $\chi^{2}$ 				&	\ChiTrESThreeKLinear$^{+\ChiPlusTrESThreeKLinear}_{-\ChiMinusTrESThreeKLinear}$										&	\ChiTrESThreeKLinearCyclicZero$^{+\ChiPlusTrESThreeKLinearCyclicZero}_{-\ChiMinusTrESThreeKLinearCyclicZero}$										& \ChiTrESThreeH$^{+\ChiPlusTrESThreeH}_{-\ChiMinusTrESThreeH}$ 									& \ChiTrESThreeHCyclicZero$^{+\ChiPlusTrESThreeHCyclicZero}_{-\ChiMinusTrESThreeHCyclicZero}$										& \ChiTrESThreeHCyclicOne$^{+\ChiPlusTrESThreeHCyclicOne}_{-\ChiMinusTrESThreeHCyclicOne}$										\\
$\Delta F$ (\%)					&	\FpOverFStarPercentAbstractTrESThreeKLinear$^{+\FpOverFStarPercentAbstractPlusTrESThreeKLinear}_{-\FpOverFStarPercentAbstractMinusTrESThreeKLinear} $	&	\FpOverFStarPercentAbstractTrESThreeKLinearCyclicZero$^{+\FpOverFStarPercentAbstractPlusTrESThreeKLinearCyclicZero}_{-\FpOverFStarPercentAbstractMinusTrESThreeKLinearCyclicZero} $	& \FpOverFStarPercentAbstractTrESThreeH$^{+\FpOverFStarPercentAbstractPlusTrESThreeH}_{-\FpOverFStarPercentAbstractMinusTrESThreeH} $	& \FpOverFStarPercentAbstractTrESThreeHCyclicZero$^{+\FpOverFStarPercentAbstractPlusTrESThreeHCyclicZero}_{-\FpOverFStarPercentAbstractMinusTrESThreeHCyclicZero}$	& \FpOverFStarPercentAbstractTrESThreeHCyclicOne$^{+\FpOverFStarPercentAbstractPlusTrESThreeHCyclicOne}_{-\FpOverFStarPercentAbstractMinusTrESThreeHCyclicOne}$		\\
3$\sigma$ upper limit on $\Delta F$ (\%)	&	$<$ \FpOverFStarPercentAbstractThreeSigmaLimitTrESThreeKLinear												&	$<$ \FpOverFStarPercentAbstractThreeSigmaLimitTrESThreeKLinearCyclicZero														& $<$ \FpOverFStarPercentAbstractThreeSigmaLimitTrESThreeH										& $<$ \FpOverFStarPercentAbstractThreeSigmaLimitTrESThreeHCyclicZero													& $<$ \FpOverFStarPercentAbstractThreeSigmaLimitTrESThreeHCyclicOne													\\
$t_{offset}$ ($min$)\tablenotemark{a}		&	\TOffsetTrESThreeKLinear$^{+\TOffsetPlusTrESThreeKLinear}_{-\TOffsetMinusTrESThreeKLinear}$								&	\TOffsetTrESThreeKLinearCyclicZero$^{+\TOffsetPlusTrESThreeKLinearCyclicZero}_{-\TOffsetMinusTrESThreeKLinearCyclicZero}$								& \TOffsetTrESThreeH \tablenotemark{b}	 												& \TOffsetTrESThreeHCyclicZero \tablenotemark{b}						 									& \TOffsetTrESThreeHCyclicOne \tablenotemark{b}																\\
$t_{eclipse}$ (HJD-2440000)			&	\JDOffsetONETrESThreeKLinear$^{+\JDOffsetPlusONETrESThreeKLinear}_{-\JDOffsetMinusONETrESThreeKLinear}$							&	\JDOffsetONETrESThreeKLinearCyclicZero$^{+\JDOffsetPlusONETrESThreeKLinearCyclicZero}_{-\JDOffsetMinusONETrESThreeKLinearCyclicZero}$							& \JDOffsetONETrESThreeH \tablenotemark{b}												& \JDOffsetONETrESThreeHCyclicZero \tablenotemark{b} 															& \JDOffsetONETrESThreeHCyclicOne \tablenotemark{b}															\\
$c_1$						&	\cOneTrESThreeKLinear$^{+\cOnePlusTrESThreeKLinear}_{-\cOneMinusTrESThreeKLinear}$									&	\cOneTrESThreeKLinearCyclicZero$^{+\cOnePlusTrESThreeKLinearCyclicZero}_{-\cOneMinusTrESThreeKLinearCyclicZero}$									& \cOneTrESThreeH$^{+\cOnePlusTrESThreeH}_{-\cOneMinusTrESThreeH}$ 									& \cOneTrESThreeHCyclicZero$^{+\cOnePlusTrESThreeHCyclicZero}_{-\cOneMinusTrESThreeHCyclicZero}$ 									& \cOneTrESThreeHCyclicOne$^{+\cOnePlusTrESThreeHCyclicOne}_{-\cOneMinusTrESThreeHCyclicOne}$										\\
$c_2$	($d^{-1}$)				&	\cTwoTrESThreeKLinear$^{+\cTwoPlusTrESThreeKLinear}_{-\cTwoMinusTrESThreeKLinear}$									&	\cTwoTrESThreeKLinearCyclicZero$^{+\cTwoPlusTrESThreeKLinearCyclicZero}_{-\cTwoMinusTrESThreeKLinearCyclicZero}$									& \cTwoTrESThreeH$^{+\cTwoPlusTrESThreeH}_{-\cTwoMinusTrESThreeH}$ 									& \cTwoTrESThreeHCyclicZero$^{+\cTwoPlusTrESThreeHCyclicZero}_{-\cTwoMinusTrESThreeHCyclicZero}$ 									& \cTwoTrESThreeHCyclicOne$^{+\cTwoPlusTrESThreeHCyclicOne}_{-\cTwoMinusTrESThreeHCyclicOne}$ 										\\
$\phi$ \tablenotemark{a}			&	\PhaseAbstractTrESThreeKLinear$^{+\PhaseAbstractPlusTrESThreeKLinear}_{-\PhaseAbstractMinusTrESThreeKLinear}$						&	\PhaseAbstractTrESThreeKLinearCyclicZero$^{+\PhaseAbstractPlusTrESThreeKLinearCyclicZero}_{-\PhaseAbstractMinusTrESThreeKLinearCyclicZero}$						& \PhaseAbstractTrESThreeH \tablenotemark{b}												& \PhaseAbstractTrESThreeHCyclicZero \tablenotemark{b}															& \PhaseAbstractTrESThreeHCyclicOne \tablenotemark{b}															\\
$e \cos(\omega)$				&	\ECosOmegaTrESThreeKLinear$^{+\ECosOmegaPlusTrESThreeKLinear}_{-\ECosOmegaMinusTrESThreeKLinear}$							&	\ECosOmegaTrESThreeKLinearCyclicZero$^{+\ECosOmegaPlusTrESThreeKLinearCyclicZero}_{-\ECosOmegaMinusTrESThreeKLinearCyclicZero}$								& \ECosOmegaTrESThreeH \tablenotemark{b} 												& \ECosOmegaTrESThreeHCyclicZero \tablenotemark{b}															& \ECosOmegaTrESThreeHCyclicOne \tablenotemark{b}															\\
% $T_{B}$	($K$)				&	\TBrightTrESThreeKLinear$^{+\TBrightPlusTrESThreeKLinear}_{-\TBrightMinusTrESThreeKLinear}$ 								&	\TBrightTrESThreeKLinearCyclicZero$^{+\TBrightPlusTrESThreeKLinearCyclicZero}_{-\TBrightMinusTrESThreeKLinearCyclicZero}$								& \TBrightTrESThreeH$^{+\TBrightPlusTrESThreeH}_{-\TBrightMinusTrESThreeH}$  								& \TBrightTrESThreeHCyclicZero$^{+\TBrightPlusTrESThreeHCyclicZero}_{-\TBrightMinusTrESThreeHCyclicZero}$								& \TBrightTrESThreeHCyclicOne$^{+\TBrightPlusTrESThreeHCyclicOne}_{-\TBrightMinusTrESThreeHCyclicOne}$									\\
$T_{B}$	($K$)					&	\TBrightTrESThreeKLinear$^{+\TBrightPlusTrESThreeKLinear}_{-\TBrightMinusTrESThreeKLinear}$ 								&	\TBrightTrESThreeKLinearCyclicZero$^{+\TBrightPlusTrESThreeKLinearCyclicZero}_{-\TBrightMinusTrESThreeKLinearCyclicZero}$								& n/a 																	& n/a																					& n/a																					\\
3$\sigma$ upper limit on $T_{B}$	($K$)	&	$<$ \TBrightThreeSigmaLimitTrESThreeKLinear														&	$<$ \TBrightThreeSigmaLimitTrESThreeKLinearCyclicZero																	& $<$ \TBrightThreeSigmaLimitTrESThreeH													& $<$ \TBrightThreeSigmaLimitTrESThreeHCyclicZero															& $<$ \TBrightThreeSigmaLimitTrESThreeHCyclicOne															\\
% $f$						& 	\fReradiationTrESThreeKLinear$^{+\fReradiationPlusTrESThreeKLinear}_{-\fReradiationMinusTrESThreeKLinear}$						&	\fReradiationTrESThreeKLinearCyclicZero$^{+\fReradiationPlusTrESThreeKLinearCyclicZero}_{-\fReradiationMinusTrESThreeKLinearCyclicZero}$						& \fReradiationTrESThreeH$^{+\fReradiationPlusTrESThreeH}_{-\fReradiationMinusTrESThreeH}$						& \fReradiationTrESThreeHCyclicZero$^{+\fReradiationPlusTrESThreeHCyclicZero}_{-\fReradiationMinusTrESThreeHCyclicZero}$						& \fReradiationTrESThreeHCyclicOne$^{+\fReradiationPlusTrESThreeHCyclicOne}_{-\fReradiationMinusTrESThreeHCyclicOne}$							\\
$f$						& 	\fReradiationTrESThreeKLinear$^{+\fReradiationPlusTrESThreeKLinear}_{-\fReradiationMinusTrESThreeKLinear}$						&	\fReradiationTrESThreeKLinearCyclicZero$^{+\fReradiationPlusTrESThreeKLinearCyclicZero}_{-\fReradiationMinusTrESThreeKLinearCyclicZero}$						& n/a																	& n/a																					& n/a																					\\
3$\sigma$ upper limit on f			& 	$<$ \fReradiationThreeSigmaLimitTrESThreeKLinear													&	$<$ \fReradiationThreeSigmaLimitTrESThreeKLinearCyclicZero																& $<$ \fReradiationThreeSigmaLimitTrESThreeH												& $<$ \fReradiationThreeSigmaLimitTrESThreeHCyclicZero															& $<$ \fReradiationThreeSigmaLimitTrESThreeHCyclicOne															\\
\enddata
\tablenotetext{a}{We account for the increased light travel-time in the system \citep{Loeb05}.}
\tablenotetext{b}{By assumption.}
\label{TableParams}
%emulateApJ change
\end{deluxetable*}
%\end{deluxetable}

Similarly to a number of our near-infrared photometric data-sets taken with CFHT/WIRCam \citep{CrollTrESTwo,CrollWASPTwelve},
our 
Ks-band photometry following the reduction exhibited an obvious background trend, $B_f$, with a 
near-linear slope. The H-band photometry also displayed a less obvious background trend.
Although these trends could be intrinsic to TrES-3, the frequency with which we find such background trends
with our other near-infrared photometric data-sets suggests that they are systematic in origin.
We thus fit both data-sets independently with a secondary eclipse model and
a linear background of the form:
\begin{equation}
B_f = 1 + c_1 + c_2 dt
\end{equation}
where $dt$ is the time interval from the beginning of the observations.
As in \citet{CrollTrESTwo} we fit for the 
best-fit secondary eclipse and background using Markov-Chain Monte Carlo methods
(\citealt{Christensen01}; \citealt{Ford}; described
for our purposes in \citealt{CrollMCMC}). We use a 5$\times$10$^{6}$ step MCMC chain. We fit
for $c_1$, $c_2$, the depth of the secondary eclipse, $\Delta F$, and the mid-eclipse phase, $\phi$.
We also quote the offset that the eclipse occurs later than the 
expected eclipse center, $t_{offset}$\footnote{we take into account the 22 $s$ delay
due to light travel-time in the system [\citealt{Loeb05}]}, as well as the best-fit mid-eclipse heliocentric (UTC) julian date, $t_{eclipse}$.
We use the \citet{Mandel02} algorithm without limb darkening to generate our best-fit secondary eclipse model. 
%We obtain our stellar and planetary parameters for TrES-3 from \citet{Torres08}.
We obtain our stellar and planetary parameters for TrES-3 from \citet{Sozzetti09}.

The results from these fits for the Ks and H-band photometry
are presented in Table \ref{TableParams}.
The best-fit secondary eclipse models are presented in Figure \ref{FigObservations}. 
The phase dependence of the best-fit secondary eclipse for our Ks-band photometry is presented in Figure \ref{FigContour}.
For our H-band photometry we are unable to detect the secondary eclipse. Thus for our H-band analysis
that follows we artificially restrict the eccentricity of TrES-3b to 0, and thus
we do not fit for the eclipse phase, $\phi$. 

To determine the
effect of systematic noise in the derived eclipse parameters
we also fit our data using the
``residual-permutation'' method \citep{Winn09}, described for our purposes
in \citet{CrollTrESTwo}.
For our H-band data we also perform 8000 iterations of a boot-strap method that 
randomly scrambles the residuals and refits the data.
The results from all these methods are presented in in Table \ref{TableParams}.
%CHANGE_HERE This is if we decide to stick with the MCMC methods
For our Ks-band data, as the ``residual permutation'' and the MCMC analyses 
result in similar unncertainties we quote the MCMC errors for the rest of the paper.
For our H-band data, the boot-strap method returns the most conservative upper limit, and
we thus quote this limit for the rest of the paper.
We also present this 3$\sigma$ upper-limit on the eclipse depth in Figure \ref{FigObservations}.

We note that for our Ks-band data specifically we also explored a quadratic expression for the background term, $B_f$:
$B_f = 1 + c_1 + c_2 dt + c_3 dt^2$,
where $c_3$ is also a fit parameter. A background of this term returned similar eclipse parameters
to that of our linear background fit,
and thus we quote our linear background MCMC fit henceforth.

%CHANGE_HERE This is if we decide to use the residual  permutation method
% Due to the excess noise for our H-band photometry, the ``residual permutation'' method
% results in slightly larger uncertainties than the MCMC method for our  H-band photometry. We quote the residual
%  permutation errors for the remainder of this paper for both bands.

\section{Discussion}

\begin{figure}
\centering
\includegraphics[scale=0.65,angle=270]{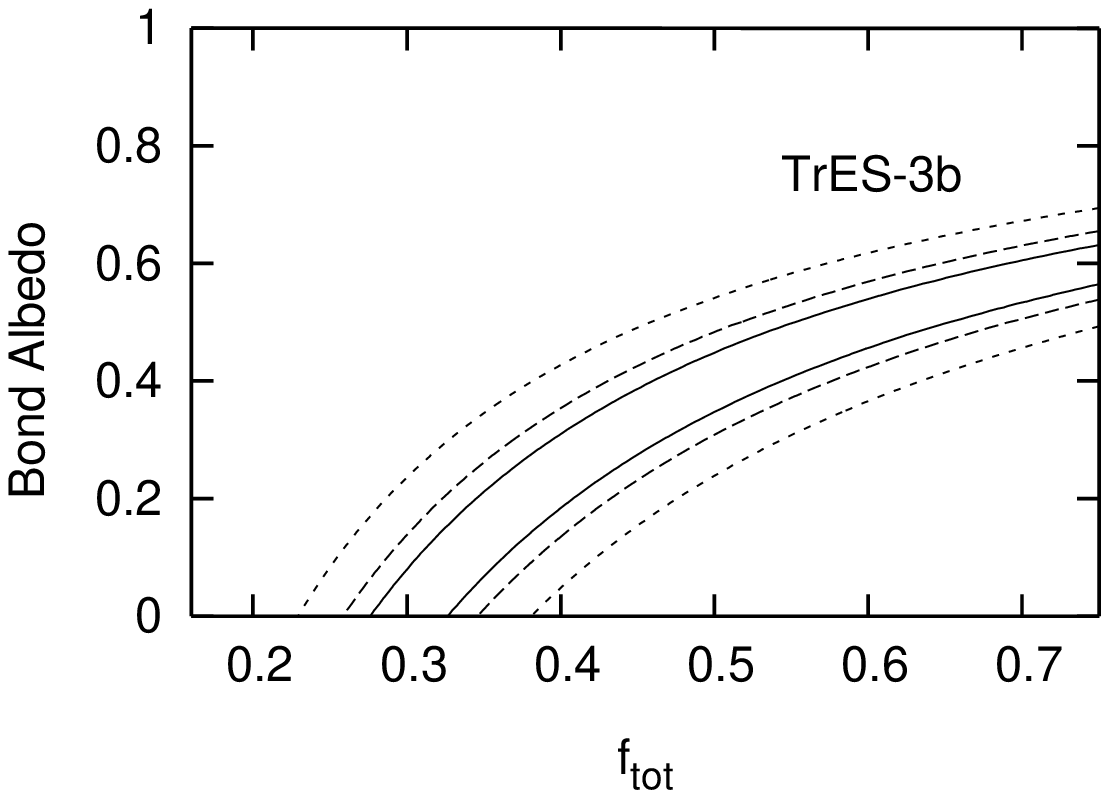}
\caption{	The 68.3\% (1$\sigma$; solid-line), 95.5\% (2$\sigma$; dashed-line)
		and 99.7\% (3$\sigma$; short dashed-line) $\chi^2$ confidence regions
		on the total reradiation factor, $f_{tot}$, and Bond albedo from
		the combination of our Ks-band point, the reanalyzed Ks-band \citet{deMooij09} point discussed here, and the
		Spitzer/IRAC measurements \citep{Fressin09}.}
\label{FigBondReradiation}
\end{figure}

 The depth of our best-fit secondary eclipse in Ks-band is 
\FpOverFStarPercentAbstractTrESThreeKLinear$^{+\FpOverFStarPercentAbstractPlusTrESThreeKLinear}_{-\FpOverFStarPercentAbstractMinusTrESThreeKLinear}$\%, with a reduced $\chi^{2}$ of \ChiTrESThreeKLinear. 
We discuss the implications of this detection combined with the other thermal emission constraints for this system in $\S$\ref{SecKband}.
In H-band we are unable to detect the secondary eclipse and discuss the implications of this in $\S$\ref{SecHband}.
We then compare these observations to atmospheric models in $\S$\ref{SecModels},
and discuss the future prospects for this system in $\S$\ref{SecFuture}.

\subsection{TrES-3b's Ks-band thermal emission}
\label{SecKband}

Our Ks-band best-fit secondary eclipse is consistent with a circular orbit;
the offset
from the expected eclipse center is: $t_{offset}$ = \TOffsetTrESThreeKLinear$^{+\TOffsetPlusTrESThreeKLinear}_{-\TOffsetMinusTrESThreeKLinear}$ minutes
(or at a phase of $\phi$=\PhaseAbstractTrESThreeKLinear$^{+\PhaseAbstractPlusTrESThreeKLinear}_{-\PhaseAbstractMinusTrESThreeKLinear}$).
This corresponds to a limit on the eccentricity and argument of periastron of
$e \cos \omega$ = \ECosOmegaTrESThreeKLinear$^{+\ECosOmegaPlusTrESThreeKLinear}_{-\ECosOmegaMinusTrESThreeKLinear}$, or a 3$\sigma$
limit of 
$|$$e$$\cos$$\omega$$|$ $<$ \ECosOmegaAbsoluteThreeSigmaLimitTrESThreeKLinear.
Our result is consistent with the more sensitive $e$cos$\omega$ limits
reported by \citet{Fressin09} from the secondary eclipse detections
at the four Spitzer/IRAC wavelengths.
Our result therefore supports the conclusion of \citet{Fressin09}
that the ``puffed-up'' radius of TrES-3b is unlikely to be due to tidal damping of the orbital eccentricity.

A secondary eclipse of 
\FpOverFStarPercentAbstractTrESThreeKLinear$^{+\FpOverFStarPercentAbstractPlusTrESThreeKLinear}_{-\FpOverFStarPercentAbstractMinusTrESThreeKLinear} $\%
corresponds to a Ks-band brightness temperature of 
$T_{BKs}$ = \TBrightTrESThreeKLinear$^{+\TBrightPlusTrESThreeKLinear}_{-\TBrightMinusTrESThreeKLinear}$ $K$ 
assuming a stellar effective temperature of $T_{eff}$ = 5650 $\pm$ 75 \citep{Sozzetti09}.
This compares to an equilibrium temperature $T_{EQ}$$\sim$1650 $K$
assuming isotropic reradiation, and a zero Bond albedo.

We should note that our Ks-band detection is discrepant from the \citet{deMooij09} Ks-band detection of 0.241 $\pm$ 0.043\%. 
%%%%%%%%%%%%
Our measurement is approximately half of their value,
and is discrepant by more than 2$\sigma$.
The best explanation for this discrepancy is the impact of systematic uncertainties for observations
in the near-infrared; \citet{deMooij09}
specifically mention several discrepant points at the beginning of their best-fit eclipse that
both increase the depth of their eclipse and lead to an eccentric eclipse center ($\phi$ = 0.4958 $\pm$ 0.0027).
If the planet is assumed to have zero eccentricity, in accordance with the Spitzer results and our own, and
these discrepant points are excluded, then the resulting best-fit eclipse is:
$\Delta F_{WHT}$ = 0.174$\pm$0.046\% (Ernst de Mooij \& Ignas Snellen, personal communication).
Thus our two measurements are consistent within 1$\sigma$ under these assumptions.
Any remaining variation between our two eclipse depths is likely statistical in nature, 
or could be due to eclipse variability between our observations and theirs. The difference in
 our two eclipse depths would necessitate a change 
in the brightness temperature of only $\sim$100$K$.

Our Ks-band secondary eclipse depth,
when combined with the secondary eclipse depths at the Spitzer/IRAC wavelengths from
\citet{Fressin09} and the \citet{deMooij09} reanalyzed eclipse depth quoted above, is consistent with a range of Bond albedos, $A_B$,
and efficiencies of the day to nightside redistribution of heat
on this presumably tidally locked planet (Figure \ref{FigBondReradiation}).
We parameterize the level of redistribution from the day to nightside by the reradiation factor,
$f$, following the \citet{Lopez07} definition, which relates the dayside
temperature of the planet, $T_p$ to the stellar effective temperature ($T_*$),
stellar radius ($R_*$), and semi-major axis of the planet ($a$): $T_p$ = $T_*$ $(R_*/a)^{1/2} [f (1-A_B)]^{1/4}$,
in the absence of any intrinsic flux (which for hot Jupiters is much smller than the absorbed and reradiated stellar flux).
From a theoretical perspective, one expects that the 
total reradiation factor, $f_{tot}$, of a hot Jupiter's atmosphere should
fall between isotropic reradiation ($f_{tot}$=$\frac{1}{4}$), and 
no redistribution ($f_{tot}$=$\frac{2}{3}$; \citealt{Burrows08b}).
For reference $f_{tot}$=$\frac{1}{2}$ denotes redistribution and reradiation from the dayside-face only.
However, the reradiation factors of individual atmospheric layers may fall well below
or above these levels, as $f$ in this case simply relates the properties of the system ($T_*$, $R_*$, and $a$)
to the brightness temperature of the atmospheric layer
probed by that wavelength of observations.
%Thus measurements of $f$ at a single wavelength
%will be directly affected by absorption and emission bands. 

 	For our Ks-band observations, if we assume a Bond albedo near zero, consistent with observations of other hot Jupiters \citep{Charbonneau99,Rowe08}
and with model predictions \citep{Burrows08},
we find a reradiation factor of $f_{Ks}$ = \fReradiationTrESThreeKLinear$^{+\fReradiationPlusTrESThreeKLinear}_{-\fReradiationMinusTrESThreeKLinear}$
from our Ks-band eclipse photometry only,
indicative of relatively efficient advection of heat from the day-to-nightside at this wavelength.
Our Ks-band reradiation factor, $f_{Ks}$, is consistent with the reradiation factor that results from combining our eclipse depth
with that of the Spitzer/IRAC depths and the \citet{deMooij09} measurement quoted above. The best-fit total reradiation factor, $f_{tot}$, 
that results from a $\chi^{2}$ analysis of all the eclipse detections for TrES-3b (and thus excluding
our H-band limit) 
assuming a zero Bond albedo is 
$f_{tot}$ = \fReradiationTrESThreeALL$^{+\fReradiationPlusTrESThreeALL}_{-\fReradiationMinusTrESThreeALL}$. 

Another way of parameterizing this redistribution is by comparing the bolometric dayside luminosity,  $L_{day}$,
of the hot Jupiter to its nightside bolometric luminosity, $L_{night}$.
Simply by following elementary thermal equilibrium calculations one can deduce that TrES-3b
should display a total bolometric luminosity of 
$L_{tot}$ = 12.5$\times$10$^{-5}$$L_{\odot}$, assuming it is in thermal equilibrium with its surroundings
and has zero Bond albedo. 
For our $f_{tot}$ = 0.301 blackbody model the dayside luminosity is 
$L_{day}$ = 7.5$\times$10$^{-5}$$L_{\odot}$, suggesting that $\sim$60\% of
the incident heat on this planet is reradiated by the dayside,
leaving $\sim$40\% to be advected to the nightside.

\subsection{An uppper-limit on TrES-3b's H-band thermal emission}
\label{SecHband}

 In H-band we are able to place a 3$\sigma$ upper-limit on 
the depth of the secondary eclipse of $\Delta F_H$ $<$ \FpOverFStarPercentAbstractThreeSigmaLimitTrESThreeHCyclicOne \%.
The 3$\sigma$ upper limit on the H-band brightness temperature is  
$T_{BH}$$<$\TBrightThreeSigmaLimitTrESThreeHCyclicOne \ $K$, a limit nearly as low as the equilibrium
temperature of TrES-3b ($T_{EQ}$$\sim$1650$K$) assuming a zero Bond albedo and isotropic reradiation.
The associated reradiation factor for the atmospheric layer probed by our H-band observations is 
$f_{H}$ $<$ \fReradiationThreeSigmaLimitTrESThreeHCyclicOne. Presuming that our limit
does not suffer from systematic effects that we have not accounted for, the H-band brightness temperature
of TrES-3 is remarkably low.

One possibility to explain the less-luminous dayside emission (weak H-band flux) of TrES-3b 
is that the albedo
of TrES-3b is significantly non-zero, as has been conjectured for a number of hot Jupiters
by \citet{CowanAgol10}. If this conjecture is true for this planet then the observed
thermal emission is not due to the planet reradiating nearly isotropically,
but due to inefficient redistribution of heat after a significant fraction of the light is reflected
(the allowed parameter space in the right-half of Figure \ref{FigBondReradiation}).
As mentioned above, a significantly non-zero albedo has been ruled out for all other hot Jupiters for which in-depth investigations
have been performed. Nonetheless TrES-3b remains an attractive target for optical, reflected light observations.

	More likely possibilites to explain the reduced H-band emission are that we are 
probing an atmospheric depth of TrES-3b that is well-homogenized, or that we are seeing a wide absorption band near
this wavelength that is depressing the observed flux. 
Simplified one-dimensional atmospheric models \citep{Fortney08} suggest that we should be seeing deeper in the atmosphere in H-band
than in Ks. Our measurements, taken at face value, imply that the atmosphere is modestly colder deeper down -- that is, we 
are seeing a small temperature inversion deep in the atmosphere (and thus at a depth much greater than the temperature inversions seen 
by others in the Spitzer/IRAC bands). Perhaps this inversion is due to efficient homogenization at high pressures
where the advectime timescale
may be of similar order to the radiative timescale
\citep{Fortney08}. Alternatively, for the absorption band possibility, there could be a strong opacity
source that is blocking our anticipated H-band opacity window, and we are actually seeing high in the atmosphere (low $P$) 
where the gas is colder. We find the former explanation compelling, but also explore the latter possibility of an additionial opacity source and a 
wide absorption band. 

One possibility for a chemical that could be causing such absorption
is methane, which has an absorption band at around 1.7 $\mu m$, which could, in principle, shave off flux from the red edge of the planet's H-band flux.
Such an absorption feature was recently detected in the emission spectrum of HD 209458 from the dayside of this planet \citep{Swain09},
and was attributed to methane.  For comparison, the L-type brown dwarfs, which reach down to
$T_{\rm eff} \sim$~1350 K \citep{Cushing08,Stephens09}, show no evidence for methane in the near-infrared, but the T-type spectral class
below 1350 K shows methane absorption in the near-infrared - indeed, that is the definition of the new spectral class.  Due to the planet's
relatively high temperatures, detectable methane in the atmosphere of TrES-3b is not expected from considerations of equilibrium
chemistry \citep{Lodders02} or non-equilibrium chemistry including vertical mixing \citep{Saumon07}.  Photochemical models of hot Jupiter atmospheres
show that methane is also readily destroyed by the incident stellar flux \citep{Zahnle09}.  However, these various chemical models are not yet
verified for hot Jupiters, so methane absorption cannot be excluded at this time.
If methane is present in a large enough quantity to suppress the H-band flux, it would significantly affect the opacities and thus the emitted flux
at other wavelengths as well, particularly at 3.3 and 8 $\mu$m, thus affecting conclusions on the efficiency of day-to-nightside redistribution and the
presence or lack thereof of a temperature inversion
for this planet.  We encourage further modelling to explore this possibility.

\subsection{Comparisons to atmospheric models}
\label{SecModels}

\begin{figure}
\centering
\includegraphics[scale=0.49,angle=270]{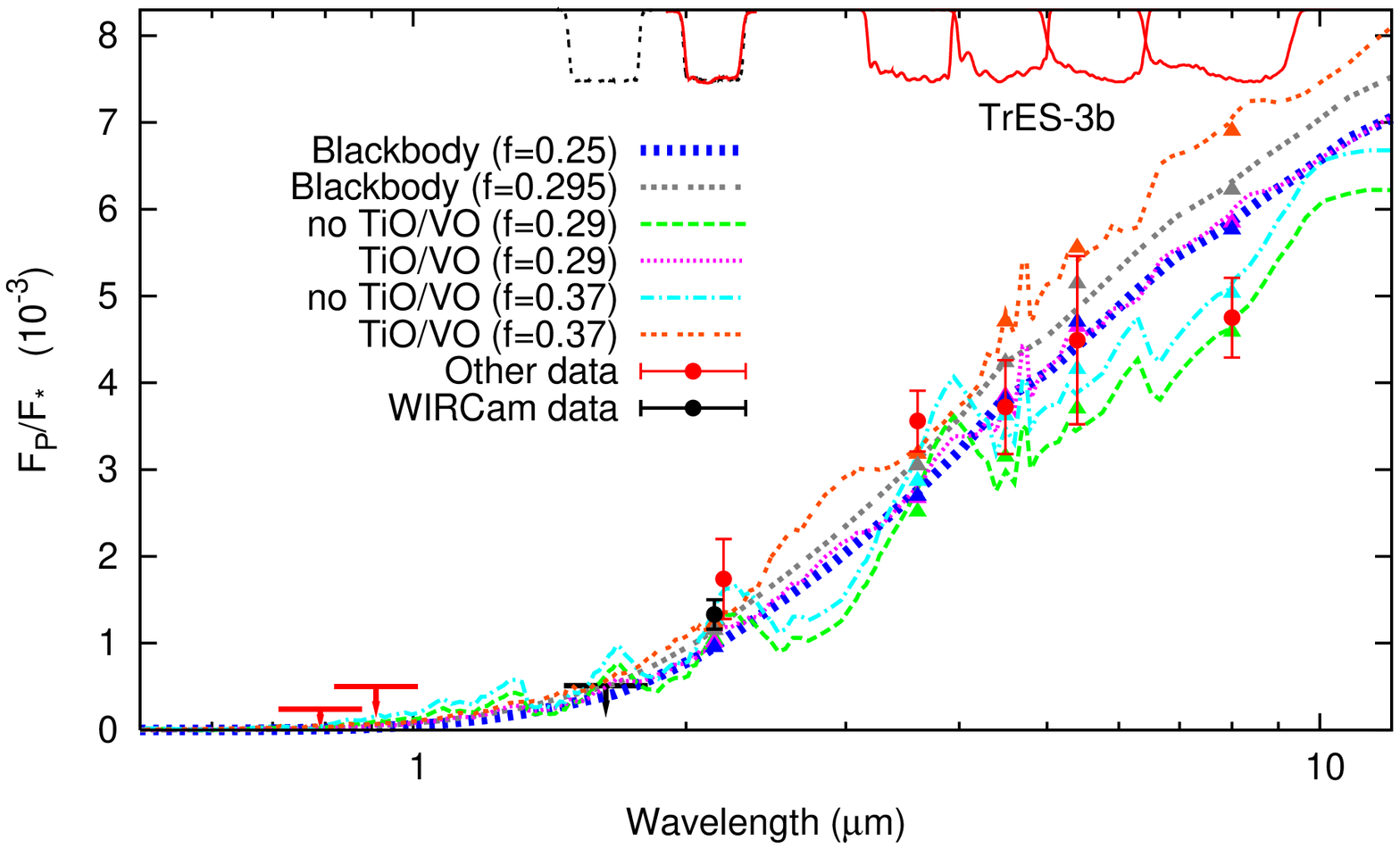}
\includegraphics[scale=0.49,angle=270]{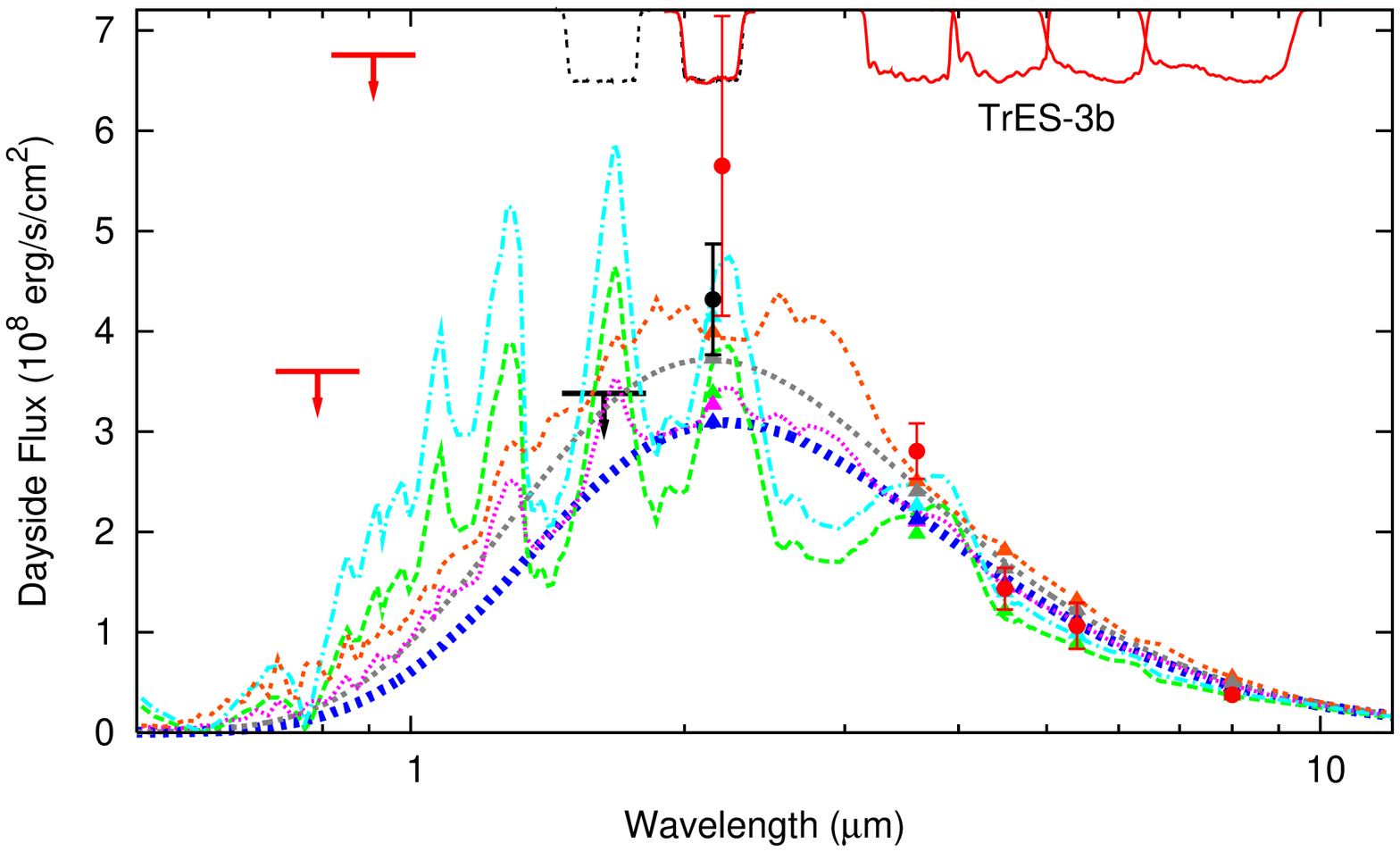}
\caption{	
		Dayside planet-to-star flux ratios (top) and dayside flux at the planet's surface (bottom).
		The Ks-band point ($\sim$2.15 $\mu m$) and H-band 3$\sigma$ upper limit (black point and black downward arrow) are our own, 
		while the red points are the Spitzer/IRAC eclipses from \citet{Fressin09}, and the WHT eclipse depth from the \citet{deMooij09} reanalyzed photometry
		as quoted above.
		We also present the 90\% upper-limits obtained by \citet{Winn08} short of 1 $\mu m$ (red downward arrows).
		Blackbody curves for isotropic reradiation ($f$=$\frac{1}{4}$; $T_{eq}$$\sim$1650 $K$; blue dashed-line)
		and our best-fit total reradiation factor ($f$=0.301; $T_{eq}$$\sim$1728 $K$; grey dotted-line) are 
		also plotted. We also plot one-dimensional, radiative transfer spectral models \citep{Fortney06,Fortney08}
		for various reradiation factors and with and without TiO/VO.
		We plot models with reradiation factors of $f$=0.37
		with and without TiO/VO (orange dotted and cyan dot-dashed lines, respectively),
		and with close to isotropic reradiation ($f$=0.29) with and without TiO/VO (magenta dotted and green dashed lines, respectively).
		Both models with TiO/VO display temperature inversions.
		The models on the top panel are divided by a stellar atmosphere model \citep{Hauschildt99} of TrES-3 
		using the parameters from \citet{Sozzetti09}
		($M_{*}$=0.928 $M_{\odot}$, $R_{*}$=0.829 $R_{\odot}$, $T_{eff}$=5650 $K$, and log $g$= 4.4).
		We plot the Ks and H-band WIRCam transmission curves (black dotted curves), as well as the Spitzer/IRAC and WHT/LIRIS
		Ks-band transmission curves (red solid curves) inverted at arbitrary scale at the top of both panels. As the WHT/LIRIS
		Ks-band transmission curve is nearly identical to the CFHT/WIRcam Ks-band transmission curve, we offset the WHT/LIRIS
		point slightly in wavelength for clarity.
	}
\label{FigModel}
\end{figure}

We compare the depth of our Ks-band eclipse, and our H-band 3$\sigma$ upper-limit on the eclipse
depth, to a series of planetary atmosphere models in Figure \ref{FigModel}.
We include the Spitzer/IRAC eclipse depths reported by \citet{Fressin09}, the revised \citet{deMooij09} eclipse depth reported above,
and the limits on thermal emission and reflected light
at shorter wavelengths of \citet{Winn08}.
This comparison is made qualitatively as well as quantitatively by integrating the models over the WIRCam H \& Ks bandpasses
as well as the Spitzer/IRAC channels and the WHT/LIRIS Ks-bandpass, and calculating the $\chi^{2}$ of the thermal emission data compared to the models.
We specifically
exclude and then include our H-band upper limit, $\chi^{2}_{NoH}$ and $\chi^{2}_{H}$, respectively, as this upper limit is
difficult to reconcile with the below models.

For the data longwards of 2 $\mu m$ the eclipse depths are relatively well-fit by blackbody
models featuring an isotropic reradiation factor ($f$=$\frac{1}{4}$; blue dotted line; $\chi^{2}_{NoH}$=\BlackbodyOneChi),
or our best-fit reradiation value ($f=0.301$; grey dotted-line; $\chi^{2}_{NoH}$=\BlackbodyTwoChi). 
These models have dayside temperatures of $T_{day}$$\sim$1650$K$ and 
$T_{day}$$\sim$1728$K$, respectively. Blackbody models that fit the 
wavelength range between $\sim$2 and $\sim$7 $\mu m$, overpredict the 8 $\mu m$ flux compared to the
Spitzer/IRAC eclipse depth at this wavelength.
We thus also compare these thermal emission measurements
to a series of one-dimensional, radiative transfer, spectral models 
\citep{Fortney05,Fortney06,Fortney08} with different reradiation factors
that specifically include or exclude gaseous TiO/VO
into the chemical equilibrium and opacity calculations.
In these models when TiO/VO are present in gaseous form in the upper atmosphere
they act as absorbers at high altitudes and lead to 
hot stratospheres and temperature inversions \citep{Hubeny03}.
% However, for temperatures lower than 1670 $K$ at 1 mbar TiO begins to condense
% and rains out of the atmosphere, and a temperature inversion will not form \citep{Fortney08}. 
% Thus below this temperature value models with and without TiO are very similar.
We present models with reradiation factors of $f$=0.29,
and $f$=0.37, both with and without TiO/VO. Both the models with TiO/VO display temperature inversions.
The brighter of these two inverted atmosphere models ($f$=0.37; orange dotted line) is a close match to our own Ks-band point,
but is generally too hot for the Spitzer wavelengths and is thus demonstrably inconsistent with the data ($\chi^{2}_{NoH}$=\FortneyOneChi).
The fainter of these two models with TiO/VO ($f$=0.29; magenta dotted line) is a reasonable fit to some wavelengths, but
is marginally discrepant from the Spitzer 8.0 $\mu m$ point ($\chi^2_{NoH}$=\FortneyTwoChi).
The models without temperature inversions are superior to those with inversions at 2.0 $\mu m$
and longer. The best-fit model is the $f$=0.37 model without TiO/VO (cyan dot-dashed line; $\chi^{2}_{NoH}$=\FortneyThreeChi), which 
provides a quantitatively and qualitatively better fit than the $f$=0.29 model (green dashed line; $\chi^{2}_{NoH}$=\FortneyFourChi). 
The differences are most obvious in the 3.6 $\mu m$ channel, but the model better predicts the other bands as well.
This thus suggests that a model without a temperature inversion with modest day-to-nightside redistribution
provides an excellent fit to the measured eclipse depths at 2.0 $\mu m$ and longwards.

However, the preceeding discussion completely ignored our strict H-band upper limit.
This is appropriate if the low level of emitted flux at this wavelength is due to an absorption band from 
a species that we do not include, or do not include at the correct concentrations, in our model atmosphere.
Nonetheless, we calculate the $\chi^{2}$ of our models including the H-band limit, $\chi^{2}_{H}$.
For reference, the predicted H-band flux from our best-fit blackbody model is 0.051\%, a value we are able to exclude
nearly at 3$\sigma$.
% while the isotropic blackbody model
%has an H-band predicted flux of 0.4\% %(different model discrepancy here)
% We assume zero-emitted flux in H-band
% rather than our brightening during secondary eclipse, and thus our limits with our 1$\sigma$ errors bars become
% $\Delta F_H$ = 0.0$^{+\FpOverFStarPercentAbstractPlusTrESThreeHCyclicOne}_{-\FpOverFStarPercentAbstractMinusTrESThreeHCyclicOne} $\%.
The $\chi^{2}$ of our former best-fit model ($f$=0.37 without TiO/VO) becomes much worse ($\chi^{2}_{H}$=\FortneyThreeChiNoH),
because our non-inverted
atmospheric model
predicts elevated flux in H-band (actually the higest flux in H-band of any of the models we present). The hottest model with
the temperature inversion ($f$=0.37 with TiO/VO) remains a poor fit ($\chi^{2}_{H}$=\FortneyOneChiNoH),
but the cooler models ($f$=0.29) with and
without TiO/VO are nearly statistically indistinguishable from our best-fit model ($\chi^{2}_{H}$=\FortneyTwoChiNoH \ 
with TiO/VO, and $\chi^{2}_{H}$=\FortneyFourChiNoH \ without TiO/VO).
Our blackbody models actually provide very similar fits to these models, as these
simple models predict much lower emission in H-band, consistent with our strict upper-limit
($\chi^{2}_{H}$=\BlackbodyOneChiNoH \ for our blackbody model with $f$=$\frac{1}{4}$, and $\chi^{2}_{H}$=\BlackbodyTwoChiNoH \
for our blackbody model with $f$=0.301).
If the pressure-temperature profile of the
atmosphere is more nearly isothermal than predicted by models, then the differences between emission peaks and troughs will be muted,
leading to more blackbody-like emission spectrum -- we find this explanation compelling for TrES-3b.

% These models have even more difficulty matching the 4.5 $\mu m$ depth that is presumably due to water and CO emission, rather than absorption, in the inverted
% atmosphere.
% If the temperature inversion is due to TiO, by the time the atmosphere becomes hot enough that TiO does not condense out, the thermal 
% radiation from the model planets is too bright to fit our own Ks-band point and the 3.6, and 5.8 $\mu m$ points.
% These models and the combination of our eclipse depths with the Spitzer/IRAC points \citep{ODonovan09}
% thus suggest that the atmosphere of TrES-2b likely features modest redistribution of heat from the day to the
% nightside, combined with a temperature inversion due to a chemical species that is predominantly not TiO.  For instance, \citet{Zahnle09} have investigated the 
% photochemistry of sulphur-bearing species as another alternative.  Given the relatively good fit of the 
% 1590 K blackbody model (solid red) to the five data points, this may indicate a
% fairly isothermal dayside temperature structure, perhaps similar to HAT-P-1b \citep{Todorov10}.

\subsection{Future prospects}
\label{SecFuture}

 Our near-infrared observations of this planet's secondary eclipse clearly show the need for multi-wavelength observations
to develop a complete understanding of the energy budgets of hot Jupiters.
 In addition, detections of secondary eclipses in multiple near-infrared
bands for multiple planets opens the door of a comparitive study of hot Jupiters and brown dwarfs at similar $T_{\rm eff}$,
to better understand how heating from above, versus heating from below, affects the temperatures and chemistry of these objects.
We will shortly reconfirm or improve upon
our H-band upper limit by observing TrES-3b in this band again, to continue to facilitate a greater understanding of this planet's 
reradiation, and advection of heat at various depths and pressures in its exotic atmosphere.

\acknowledgements

The Natural Sciences and Engineering Research Council of Canada supports the research of B.C. and R.J.
The authors would like to thank Ernst de Mooij and Ignas Snellen for helpful discussions and for refitting their observations.
The authors would like to thank Marten van Kerkwijk for helping to optimize these observations.
The authors especially appreciate the hard-work and diligence of the CFHT staff
in helping us pioneer this ``stare'' method on WIRCam. We thank the anonymous referee for a thorough review.

% \appendix
% \begin{figure*}
% \centering
% \includegraphics[scale=0.45,angle=90]{Planet_K.ps}
% \includegraphics[scale=0.45,angle=90]{Planet_H.ps}
% \caption{	Left panel: The CFHT/WIRCam full frame array during our Ks-band 
% 		observations of TrES-3b. The image has been preprocessed
% 		with the `I`iwi pipeline; the obvious artifacts (in the lower-left corners of the bottom two chips
% 		for instance or the crosses on the upper two chips) are due to the masking of bad pixels.
% 		The target star, TrES-3 (large square), and the reference stars 
% 		used to correct the flux of TrES-3 (circles) are marked.
% 		Right panel: The same for our H-band observations.}
% \label{FigTrES3K}
% \end{figure*}

\end{document}